\documentclass[preprint,twocolumn]{article}
\usepackage{amsmath}
\usepackage{amsfonts}
\usepackage{graphicx}
\usepackage{xcolor}
\usepackage{multicol,caption}
\usepackage{caption}
\usepackage[hyphens]{url}
\usepackage{hyperref}
\usepackage[hyphenbreaks]{breakurl}

\hypersetup{breaklinks=true}

\usepackage{cite}

\begin{document}

\title{Hall transport in the topological non-Hermitian checkerboard lattice}
\author{Pedro G. de Oliveira$^a$\thanks{Email: pgo.pedro@gmail.com} \, and Antônio S. T. Pires$^a$}
\date{\today}

\twocolumn[
  \begin{@twocolumnfalse}
    \maketitle
    \begin{center}
	\textit{$^a$Department of Physics, Federal University of Minas Gerais, Belo Horizonte, MG, Brazil.} \newline  		
	\textit{$^*$pgo.pedro@gmail.com} 
	\end{center}
	\vspace{\baselineskip}

    \begin{abstract}
		The checkerboard lattice is a two-dimensional non-trivial structure usually seen as a planar version of the pyrochlore lattice. This geometry supports a two-band insulating electronic system with Chern topology induced by a complex hopping parameter. Inspired by the recent advances in the topology of non-Hermitian systems, in this work we study a non-Hermitian version of the topological checkerboard lattice. The complex band structure and Berry curvature are calculated. In the insulating phase, the Chern number is the same as in the Hermitian version, but the Hall conductivity is no longer quantized. The dependence of the Hall conductivity with the non-Hermitian parameters is investigated. The non-Hermiticity can be seen as a result of dissipation caused by coupling the system to the environment, so this study casts light on the topology of open systems in condensed matter physics.

    \end{abstract}
    
\vspace{\baselineskip}
\vspace{\baselineskip}

  \end{@twocolumnfalse}
]

\section{Introduction}

The study of condensed matter systems from a topological point of
view has elucidated interesting novel phenomena in the last decades, the most famous example being the concept of topological insulators \cite{Hasan2010,Fruchart2013}. The topological approach was fruitful in a variety of models, such as
magnetic, photonic, acoustic and superconducting systems \cite{Ryu2010,Qi2011,Ssstrunk2015,Goldman2016,Ozawa2019,Malki2020}.
While topological effects in regular, energy-conserving Hamiltonians
is a established fact with robust theoretical and experimental evidence,
that investigation in non-conservative open systems is still in progress.
One way to model an open system is with a non-Hermitian Hamiltonian.
The non-Hermitian formalism can be applied to a variety of quantum
and classical systems, and the topology of non-Hermitian systems is
an ongoing field of study \cite{Ashida2020,Tlusty2021,Hurst2022,Banerjee2023,Yu2024}.
Specifically in fermionic systems, it is known that the non-Hermiticity
spoils the bulk-edge correspondence of Chern topological insulators,
and some efforts have been made to restore it \cite{Yao2018}. Another
consequence is that the Hall conductivity is no longer quantized \cite{Hirschberger2015,Chen2018_PRB98,Hirsbrunner2019_PRB100}.
For that reason, exploring the interplay between non-Hermiticity and
topological effects on different platforms is a valuable way to elucidate
the physics of open systems.

The checkerboard lattice is a nontrivial two-dimensional lattice structure
with square geometry, which can be seen as a planar version of the
pyrochlore lattice \cite{Canals2002,Fujimoto2002,Bernier2004,Pollmann2006,Yoshioka2008,Santos_2010,Sun2009,Sun2011,Katsura2010_linegraph,Liu2022_linegraphs,Liu2017,Pires2019,Pires2020,Pires2021_2,Ma2020,Zhang2021}.
The experimental realization is challenging, but the structure has
already been discovered in some compounds \cite{Kabbour2008,Hu2023,Sufyan2024}.
The checkerboard lattice has some interesting properties. For instance,
it is a member of a set of structures known as line-graph lattices,
which show flat bands and spatially localized states \cite{Katsura2010_linegraph,Liu2022_linegraphs}.
It has also been shown that it can hold variety of topological states
\cite{Canals2002,Bernier2004,Katsura2010_linegraph,Santos_2010,Sun2009,Sun2011,Liu2017,Pires2019,Pires2020,Pires2021_2,Ma2020,Zhang2021}.
One interesting context to study these topological states is in a
dissipative environment, which can be represented by a non-Hermitian
Hamiltonian.

Motivated by the nontrivial topology of the fermionic checkerboard
lattice, the main purpose of this work is to study a non-Hermitian
version of this lattice, with a focus on Hall transport and its
response to the non-Hermitian parameters. This paper is organized
as follows: In Section 2 we discuss the description of open quantum
systems with non-Hermitian Hamiltonians. In Section 3 we introduce
the checkerboard lattice and show results for the fermionic Hermitian
Hamiltonian. In Section 3 we introduce the non-Hermitian Hamiltonian
and calculate its complex band structure. In Section 5 we calculate
the system's Hall conductivity and show how it varies in response
to the non-Hermitian parameters. In Section 6 we summarize our findings. 

\section{Non-Hermitian Hamiltonians for open systems}

Non-Hermitian Hamiltonians are a well-known way to describe open quantum
systems, which present mechanisms of gain and loss of energy or particles.
These open systems account for many physical processes, for instance,
the interaction of a quantum system with a surrounding environment
or an experimental apparatus of continuous measures \cite{Ashida2020}.
Regarding the interaction of a two-band quantum system with its surroundings,
one way to describe it is to locally couple the system to a series
of baths (which represent the environment) \cite{Midtgaard2019}.
That formalism accounts for a lossy system, where quasiparticles have
finite lifetime due to dissipation. Integrating out the baths' degrees
of freedom we obtain a Green function of the form $G\left(\mathbf{k},\epsilon\right)=\left(\epsilon-H\left(\mathbf{k}\right)-\Sigma\left(\mathbf{k},\epsilon\right)\right)^{-1}$.
The term $\Sigma\left(\mathbf{k},\epsilon\right)$ is the self-energy
and contains information about the bath modes. The effective Hamiltonian
$H_{eff}\left(\mathbf{k}\right)$ contains a dissipation term $H_{diss}=i\left(\gamma_{i}\sigma_{i}\right),$
$i=\left(0,x,y,z\right)$, which makes it non-Hermitian. Non-Hermitian
Hamiltonians have complex eigenvalues and distinct right and left
eigenstates, demanding the use of a biorthogonal algebra \cite{Brody2013,Groenendijk2021}.
In the formalism of coupled baths above, the physical requirement
that the system's density of states is non-negative results in a constraint
to the complex eigenvalues: their imaginary part has to be non-positive\cite{Midtgaard2019}.

Regarding the transport properties of quasiparticles, it is established
that the Hall conductivity for a two-dimensional fermionic system
can be obtained from the Matsubara Green function $G$ as \cite{Ishikawa1986,Ishikawa1987,Hirsbrunner2019_PRB100}:
\begin{equation}
\sigma_{xy}=\frac{e^{2}}{h}\frac{\varepsilon_{\mu\nu\rho}}{24\pi^{2}}\int d^{3}p\,Tr\left[G\frac{\partial G}{\partial p_{\mu}}G\frac{\partial G}{\partial p_{\nu}}G\frac{\partial G}{\partial p_{\rho}}\right]\label{eq:conduc_general}
\end{equation}

where we assume $T=0\:K$ so the sum over Matsubara frequencies becomes
an integral. The expression above works for either Hermitians and
non-Hermitians Hamiltonians, provided that the Green function is known.
For a Hermitian Hamiltonian, this Hall conductivity is quantized and
gives rise to the integer Hall effect. That is a consequence of the
non-trivial topology of the Hamiltonian's Bloch bundle \cite{Hasan2010,Fruchart2013}.
The quantization is indexed by the Chern number, or TKNN integer \cite{TKNN}.
The system is a topological insulator due to the bulk-edge correspondence.
However, for a non-Hermitian system, the self-energy $\Sigma\left(\mathbf{k},\epsilon\right)$
spoils this quantization even if the system has non-trivial Chern
topology \cite{Hirschberger2015,Hirsbrunner2019_PRB100}. In this
work, this is shown explicitly in the case of the non-Hermitian checkerboard
lattice. 

\section{Hermitian checkerboard lattice}

\begin{figure}[t!]
\center
\includegraphics[scale=0.7]{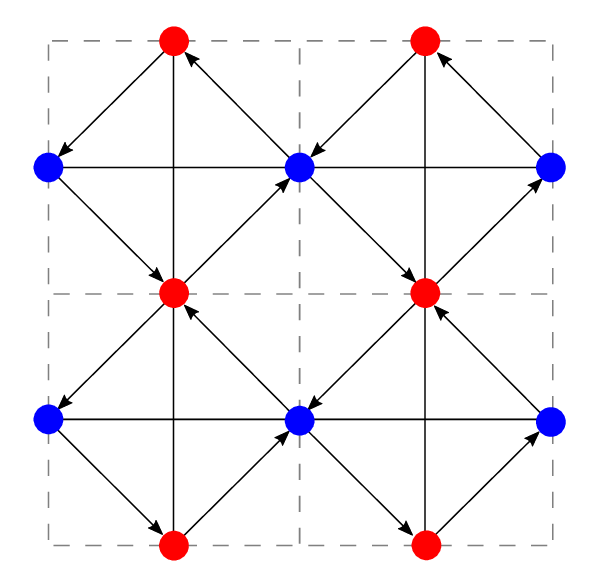}
\caption{The checkerboard lattice. The inequivalent sites are represented in
blue and red. The NN hopping is complex, with a phase of $+\phi$
in the direction of the arrows and $-\phi$ against it. The NNN hopping
is represent by the solid vertical and horizontal lines. We chose
the lattice parameter as $a=1$. \label{fig:crystal}}
\end{figure}

To investigate the Hall conductivity of the checkerboard fermionic
lattice, we begin by analyzing the Hermitian case. The checkerboard
lattice is represented in Figure \ref{fig:crystal}, with tight-binding
Hamiltonian written as:

\begin{align}
H^{H} & =-t\sum_{\langle i,j\rangle}e^{i\phi_{ij}}\left(c_{i}^{\dagger}c_{j}+h.c.\right)-\sum_{\langle\langle i,j\rangle\rangle}t'_{ij}\left(c_{i}^{\dagger}c_{j}+h.c.\right).\label{eq:hamilt}
\end{align}

Here, $c_{i}^{\dagger}\left(c_{i}\right)$ is the fermion creation
(annihilation) operator at site $i$. The pairs $\langle i,j\rangle$
and $\langle\langle i,j\rangle\rangle$ represent near-neighbor (NN)
and next-near-neighbor (NNN) sites, respectively. There are two inequivalent
sites in the primitive cell. The NN hopping is complex, carrying a
phase factor $\phi_{ij}=\pm\phi$ with the sign determined by the
arrows' directions in Figure \ref{fig:crystal}. That term breaks
the time-reversal symmetry for $\phi\neq n\pi$ ($n\in\mathbb{Z}$)
and allows non-trivial topology \cite{Sun2011}. The NNN hopping $t'_{ij}=t'_{1}\left(t'_{2}\right)$
is represented by solid (dashed) lines. As a minimal model, we take
$t'_{2}=0$ and define $t' \equiv t'_{1}$. In the momentum space we have a
two-band system, and the general Hermitian Hamiltonian is:
\begin{equation}
H^{H}=\sum_{k}\psi_{k}^{\dagger}H_{k}^{H}\psi_{k}
\end{equation}

with $\psi^{\dagger}=\left(\begin{array}{cc}
a_{k}^{\dagger} & b_{k}^{\dagger}\end{array}\right)$. The checkerboard Hamiltonian matrix takes the form

\begin{equation}
H_{k}^{H}=d_{0}I+d_{x}\sigma_{x}+d_{y}\sigma_{y}+d_{z}\sigma_{z}\label{eq:hamilt_hermitian}
\end{equation}

with

\begin{align}
d_{0} & =-t'\left(cos\,k_{x}+cos\,k_{y}\right)\nonumber \\
d_{x} & =-4t\,cos\,\phi\left(cos\,\frac{k_{x}}{2}\,cos\,\frac{k_{y}}{2}\right)\nonumber \\
d_{y} & =-4t\,sin\,\phi\left(sin\,\frac{k_{x}}{2}\,sin\,\frac{k_{y}}{2}\right)\nonumber \\
d_{z} & =-t'\left(cos\,k_{x}-cos\,k_{y}\right).\label{eq:parameters_hemitian}
\end{align}

The Hamiltonian's eigenvalues are
\begin{equation}
\epsilon_{\pm}=d_{0}\pm d,\qquad d\equiv\sqrt{d_{x}^{2}+d_{y}^{2}+d_{z}^{2}}
\end{equation}

and the spectrum is gapped for $\phi\neq m\pi$ ($m\in\mathbb{Z}/2$). 

\begin{figure*}[t!]
\center
\includegraphics[scale=0.65]{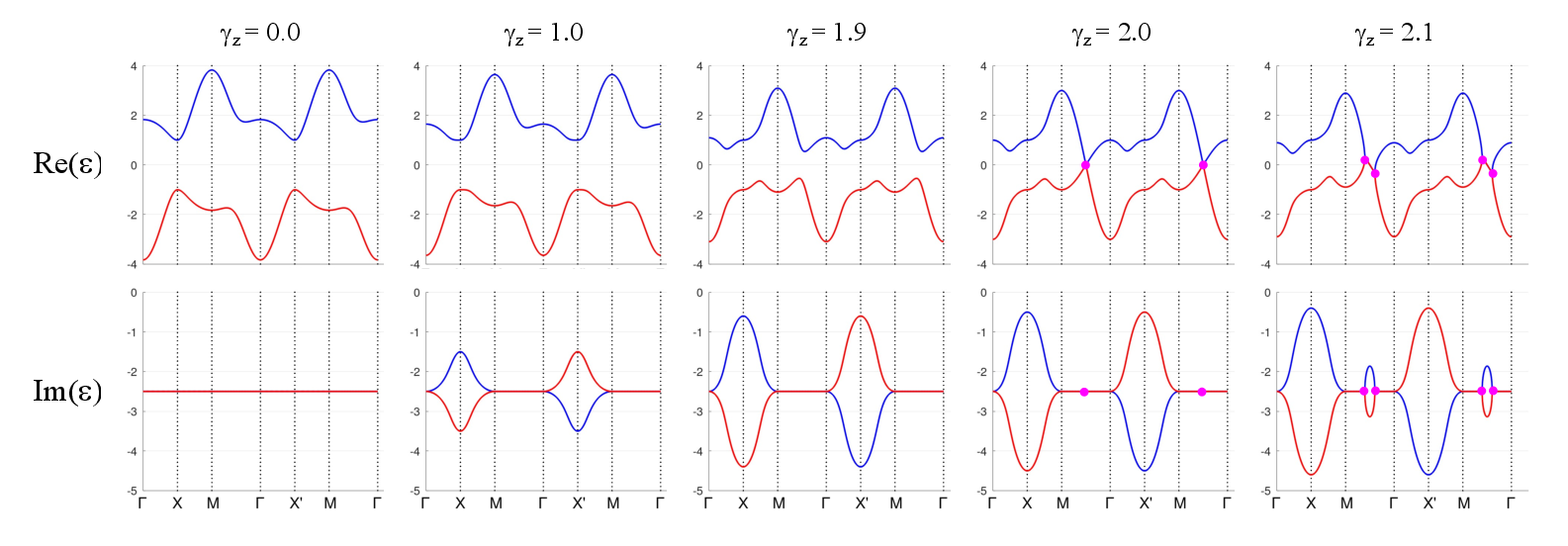}

\caption{Energy spectrum (real and imaginary parts) of the system for different
values of $\gamma_{z}$. The points of high symmetry are $\Gamma (0,0), \, X(\pi,0), \, X'(0,\pi), \, M(\pi,\pi)$. For $\gamma_{z}=2$, a hybrid
exceptional point (in magenta) is generated in the center of the $\Gamma-M$
line, which splits in two ordinary EPs for $\gamma_{z}>2$. In this work, we focus in the $\gamma_{z}<2$ regime, where
the bands do not touch and the Berry curvature is well-defined. The
other parameters of the theory are $t=1.0$, $t'=0.5$, $\phi=\pi/4$,
$\gamma_{0}=-2.5.$\label{fig:spectrum}}
\end{figure*}

The Berry curvature of the $n\text{-th}$ band is a local function of the momenta constructed from the eigenstates and defined as $\Omega_{n,ij}\left(\mathbf{k}\right)=i \varepsilon_{ij} \left\langle \partial_{i}\psi_{n}\left(\mathbf{k}\right)\middle|\partial_{j}\psi_{n}\left(\mathbf{k}\right)\right\rangle $ (where $\partial _i \equiv \frac{\partial}{\partial k_i}$).
For a two-band system with Hamiltonian written as Eq. \ref{eq:hamilt_hermitian},
the Berry curvature of the valence band can be written as \cite{Hasan2010,Fruchart2013,Pires_topology}:

\begin{equation}
\Omega_{xy}\left(\mathbf{k}\right)=\hat{\mathbf{d}}\cdot\left(\partial_{k_x}\hat{\mathbf{d}}\times\partial_{k_y}\hat{\mathbf{d}}\right)\label{eq:berry}
\end{equation}

with $\hat{\mathbf{d}}=\mathbf{d}/d$, $\mathbf{d}=\left(d_{x},d_{y},d_{z}\right).$
From the linear response theory, the Hall conductivity is given by: 

\begin{equation}
\sigma_{xy}=-\frac{e^{2}}{h}\int\frac{d^{2}k}{4\pi}\Omega_{xy}\left(\mathbf{k}\right)=-\frac{e^{2}}{h}C.\label{eq:sigma_quantized}
\end{equation}

Here, $C \equiv \int\frac{d^{2}k}{4\pi}\Omega_{xy}\left(\mathbf{k}\right)$
is the Chern number. Hence, the Hall conductivity is quantized, realizing
the integer quantum Hall effect. In the case of a two-band insulator,
the Chern number has an natural geometrical interpretation. As $\mathbf{k}$
spreads over the Brillouin torus, the parameter vector $\mathbf{d}$
describes a surface. The Chern number represents the number of times
this surface wraps around the origin (the winding number).

In Hermitian systems, a non-trivial Chern number is related to robust
conducting edge states via bulk-boundary correspondence, making the
system a topological insulator. In the gapped checkerboard lattice,
the Chern number is $C=-1$ if $\phi$ lies in the first or third
quadrant, and $C=+1$ for $\phi$ in the second or fourth quadrant.
For $\phi=n\pi/2$, the system is ungapped. The bands touch quadratically
at the corners of the Brillouin zone for even $n$, and in the center
for odd $n$.

\section{Non-Hermitian case}

One way of obtaining a non-Hermitian Hamiltonian is to include an
anti-Hermitian diagonal term in the momentum-space Hamiltonian:
\begin{align}
H^{AH} & =\sum_{k}\psi_{k}^{\dagger}H_{k}^{AH}\psi_{k}
\end{align}

where
\begin{align}
H_{k}^{AH} & =i\left(\begin{array}{cc}
\gamma_{a} & 0\\
0 & -\gamma_{b}
\end{array}\right) = i\left(\begin{array}{cc}
\gamma_{0}+\gamma_{z} & 0\\
0 & \gamma_{0}-\gamma_{z}
\end{array}\right)\nonumber \\
 & =i\left(\gamma_{0}I+\gamma_{z}\sigma_{z}\right).
\end{align}

Here, we defined $\gamma_{0}\equiv\frac{\left(\gamma_{a}-\gamma_{b}\right)}{2}$
and $\gamma_{z}\equiv\frac{\left(\gamma_{a}+\gamma_{b}\right)}{2}$.

The total non-Hermitian Hamiltonian $H_{k}^{NH}=H_{k}^{H}+H_{k}^{AH}$
can be written with complex parameters $h_{i}\equiv d_{i}+i\gamma_{i}$
as
\begin{equation}
H_{k}=h_{0}I+h_{x}\sigma_{x}+h_{y}\sigma_{y}+h_{z}\sigma_{z}.
\end{equation}

This is a general form. In our system, the only non-real parameters
(which carry the non-Hermiticity) are $h_{0}=d_{0}+i\gamma_{0}$ and
$h_{z}=d_{z}+i\gamma_{z}$. Diagonalizing the Hamiltonian we obtain
the eigenvalues:

\begin{align}
\epsilon_{\pm} & =h_{0}\pm h
\end{align}

with

\begin{align}
h & =\sqrt{h_{x}^{2}+h_{y}^{2}+h_{z}^{2}}=\sqrt{d_{x}^{2}+d_{y}^{2}+\left(d_{z}+i\gamma_{z}\right)^{2}} \nonumber \\
  & =h^{R}+ih^{I}. \label{eq:complex}
\end{align}

The eigenvalues are complex, with:
\begin{align}
Re\:\epsilon_{\pm} & =d_{0}\pm h^{R}\nonumber \\
Im\:\epsilon_{\pm} & =\gamma_{0}\pm h^{I}.
\end{align}

Within the interpretation of system-environment coupling derived in
Ref. \cite{Midtgaard2019}, it is imperative that imaginary part be
negative, ensuring the density of states is non-negative. To fulfill
this property it is sufficient to assume $\gamma_{0}<0$ and $\left|\gamma_{0}\right|>\gamma_{z}$.
The spectrum is shown in Fig. \ref{fig:spectrum} for different values
of $\gamma_{z}.$ For high values of $\gamma_{z}$ the gap closes
in exceptional points (EPs). Exceptional points are a unique and interesting
characteristic of non-Hermitian systems. In an EP, in addition to
the gap closing $\epsilon_{+}=\epsilon_{-}$, the eigenvalues of the
system coalesce and the Hamiltonian is said to be defective \cite{Ashida2020}.
For our system, Fig. \ref{fig:spectrum} shows hybrid exceptional
points at $\mathbf{k}_{HEP}=\left(\pm\pi/2,\pm\pi/2\right)$ when
$\gamma_{z}=2$. Those are called hybrid because they result from
the merging of two ordinary exceptional points with opposite vorticities
or winding numbers \cite{Shen2018,Zhang2018_HEP,Jin2020}. In fact,
for $\gamma_{z}>2$, each hybrid EP splits in two ordinary EPs localized
in the $\Gamma-M$ line. These coupled EPs are linked by a path in
which the eingenvalues of both bands are complex-conjugate of each
other, which is sometimes called bulk Fermi arcs \cite{Zhou2018,Tlusty2021}
and are a consequence of $\mathcal{PT}$ symmetry of the Hamiltonian
\cite{Hurst2022}. 

Another unique feature of non-Hermitian systems is the topological
classification derived from the vorticity of EPs, which is related
to the energy eigenvalues \cite{Shen2018,Dey2024} (rather than the
eigenstates like in Chern or $\mathbb{Z}_{2}$ insulators \cite{Hasan2010}).
This classification is not the focus of the present work and is left
for forthcoming studies. For the investigation of transverse transport,
we restrict ourselves to the gapped regime.

\section{Hall Conductivity}

\begin{figure*}[ht]
\center
\includegraphics[scale=0.7]{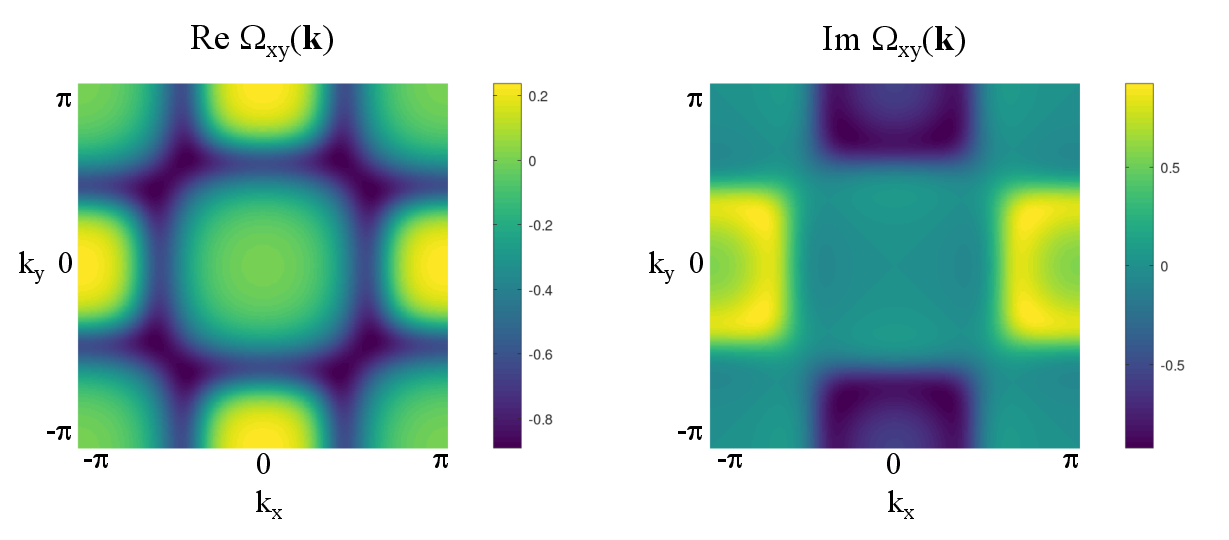}
\caption{Berry curvature of the non-Hermitian checkerboard lattice. Parameters:
$t=1.0$, $t'=0.5$, $\phi=\pi/4$, $\gamma_{z}=1.5$.\label{fig:berry} }
\end{figure*}

We can generalize the Berry curvature in Eq. \ref{eq:berry} to non-Hermitian
systems, obtaining a non-Hermitian complex Berry curvature \cite{Chen2018_PRB98}:

\begin{equation}
\Omega_{xy}\left(\mathbf{k}\right)=\hat{\mathbf{h}}\cdot\left(\partial_{k_x}\hat{\mathbf{h}}\times\partial_{k_y}\hat{\mathbf{h}}\right)\label{eq:berry_nh}
\end{equation}

with $\hat{\mathbf{h}}=\mathbf{h}/h$, $\mathbf{h}=\left(h_{x},h_{y},h_{z}\right).$
Actually, because of the difference between right and left eigenstates,
we can construct four different ``Berry curvatures''

\begin{equation}
\Omega_{xy}^{\alpha\beta}\left(\mathbf{k}\right)=i \varepsilon_{xy}\left\langle \partial_{k_x}\psi^{\alpha}\left(\mathbf{k}\right)\middle|\partial_{k_y}\psi^{\beta}\left(\mathbf{k}\right)\right\rangle,
\end{equation}

where $\alpha,\beta=L/R$ \cite{Shen2018}. We use here $\Omega_{xy}\left(\mathbf{k}\right)=\Omega_{xy}^{LR}\left(\mathbf{k}\right)$,
which is well-suited for systems with loss and gain. The non-Hermitian
Chern number is defined as in the Hermitian case: $C\equiv\int\frac{d^{2}k}{4\pi}\Omega_{xy}^{\alpha\beta}\left(\mathbf{k}\right)$.
It is a real and quantized number and does not depend
on $\alpha,\beta$, even though $\Omega_{xy}^{\alpha\beta}\left(\mathbf{k}\right)$
are locally different.

In ordinary Hermitian systems, non-trivial Chern numbers are related
to robust edge states through the bulk-edge correspondence, characterizing
a topological insulator. This bulk-edge correspondence breaks down
in non-Hermitian systems. It has been argued that it can be restored
defining a non-Bloch Chern number as the integral of the Berry curvature over
an extended complex Brillouin zone \cite{Yao2018}. Nevertheless,
the bulk Hall transport is unaffected by that. The Hall conductivity
in Eq. \ref{eq:conduc_general} can be written as \cite{Hirsbrunner2019_PRB100}:

\begin{align}
\sigma_{xy} & =-\frac{e^{2}}{h}\int\frac{d^{2}k}{2\pi^{2}}Re\left[\hat{\mathbf{h}}\cdot\left(\partial_{k_x}\hat{\mathbf{h}}\times\partial_{k_y}\hat{\mathbf{h}}\right) \times \right. \nonumber \\
& \times \left. \left(\frac{\pi}{2}sgn\left(Re\,h\right)-\frac{ihh_{0}}{h^{2}-h_{0}^{2}}-i\,arctanh\left(\frac{h_{0}}{h}\right)\right)\right]\nonumber \\
 & =-\frac{e^{2}}{h}\int\frac{d^{2}k}{4\pi}Re\left[\Omega_{xy}\left(\mathbf{k}\right)f\left(\mathbf{k}\right)\right].
\end{align}

The conductivity now bears a factor $f\left(\mathbf{k}\right)$ in
the integrand, and we cannot identify the integral with the Chern
number. Therefore, there is no topological interpretation of the Hall
response, and the Hall conductivity for a non-Hermitian topological
Chern insulator is in general non-quantized. Taking the Hermitian
limit $\mathbf{h}\rightarrow\mathbf{d}$, we can show that the expression
above reduces to Eq. \ref{eq:sigma_quantized}, restoring the integer
quantum Hall effect.

In the insulating regime, the non-Hermitian Chern number was numerically
calculated and is equal to $C=\pm1$, like in the Hermitian system.
The values of $\phi$ for which the gap closes are not only $n\pi/2$
(as in the Hermitian case), but an interval around those values. In
that interval, the gap closes in exceptional points, the Berry curvature
diverges and the Chern number is not defined. As long as we stay in
the insulating phase, the Chern topology is identical to the Hermitian
system. The real and imaginary parts of the Berry curvature are plotted
in Figure \ref{fig:berry}. Note that $Im \, \Omega_{xy}\left(\mathbf{k}\right)$
is an odd function ($Im\:\Omega_{xy}\left(-\mathbf{k}\right)=-Im\:\Omega_{xy}\left(\mathbf{k}\right)$),
so its integral is null, making the Chern number a real integer.

\begin{figure}[t!]
\center
\includegraphics[scale=0.3]{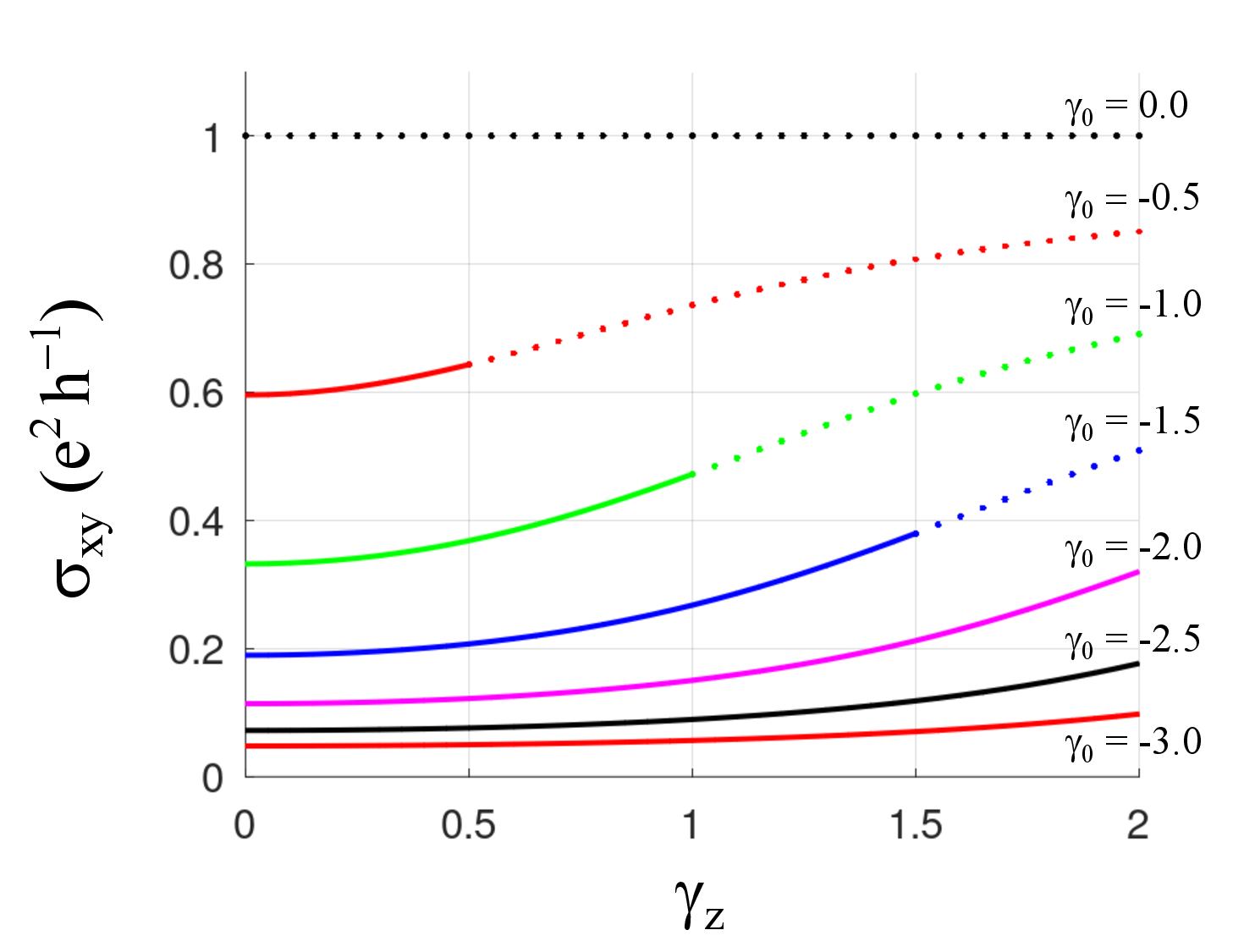}

\caption{Hall conductivity of the non-Hermitian checkerboard lattice as a function
of $\gamma_{z}$. Different curves represent different values of $\gamma_{0}.$
The other parameters are $t=1.0$, $t'=0.5$ and $\phi=\pi/4$. The
dotted sections of the plots are not physical in the coupled baths
model ($\gamma_{z}>\left|\gamma_{0}\right|$). The data does not go
further than $\gamma_{z}=2.0$ because this is the value of gap closing,
when the Berry curvature is divergent.\label{fig:Hall}}
\end{figure}

The Hall conductivity as a function of $\gamma_{z}$ is plotted in
Figure \ref{fig:Hall}. The conductivity increases with the raise
of $\gamma_{z}$, but decreases with the raise of $\left|\gamma_{0}\right|$.
This is in accordance with the interpretation that, for systems coupled
to the environment, $\left|\gamma_{0}\right|>\gamma_{z}$ represent
a net loss, i.e., particles with finite lifetimes \cite{Groenendijk2021}.
A high value of $\left|\gamma_{0}\right|$ means higher loss (or shorter
quasiparticle lifetime), decreasing the charge carriers in equilibrium
and damping the conductivity. The opposite occurs for a high $\gamma_{z}$,
which means longer quasiparticle lifetime (less loss), resulting in
a higher conductivity.

\section{Summary and concluding remarks}

We studied the fermionic non-Hermitian checkerboard lattice with focus
on its Chern topology and Hall conductivity. The non-Hermiticity is
introduced by two parameters: $\gamma_{0},$ related to dissipation,
and $\gamma_{z}$, related to gain. A complex NN hopping induces a
Berry curvature, which renders the system a non-trivial Chern topology
in the insulating phase (identically to the Hermitian case). The gap
closes in exceptional points for some choices of parameters, in specific
for high $\gamma_{z}$. Although the system is a Chern insulator, the Hall conductivity
is not quantized. This is a consequence of
dissipation or finite lifetime of states in a context of a system
coupled to the environment. We verified the response of the Hall conductivity
to the non-Hermitian parameters, and concluded it is maximized for
high $\gamma_{z}$ and low $\gamma_{0},$ which represents long quasiparticle
lifetimes. Our work explores the novel field of non-Hermitian topology and corroborates that two-dimensional exotic lattices, like the checkerboard lattice, can be a good platform to investigate topological effects in open quantum systems.

\bibliographystyle{elsarticle-num} 
   	\bibliography{refs_database}

\begin{thebibliography}{10}
\expandafter\ifx\csname url\endcsname\relax
  \def\url#1{\texttt{#1}}\fi
\expandafter\ifx\csname urlprefix\endcsname\relax\def\urlprefix{URL }\fi
\expandafter\ifx\csname href\endcsname\relax
  \def\href#1#2{#2} \def\path#1{#1}\fi

\bibitem{Hasan2010}
M.~Z. Hasan, C.~L. Kane,
  \href{http://dx.doi.org/10.1103/RevModPhys.82.3045}{Colloquium: Topological
  insulators}, Rev. Mod. Phys. 82~(4) (2010) 3045–3067.
\newblock \href {https://doi.org/10.1103/revmodphys.82.3045}
  {\path{doi:10.1103/revmodphys.82.3045}}.
\newline\urlprefix\url{http://dx.doi.org/10.1103/RevModPhys.82.3045}

\bibitem{Fruchart2013}
M.~Fruchart, D.~Carpentier,
  \href{http://dx.doi.org/10.1016/j.crhy.2013.09.013}{An introduction to
  topological insulators}, C. R. Phys. 14~(9–10) (2013) 779–815.
\newblock \href {https://doi.org/10.1016/j.crhy.2013.09.013}
  {\path{doi:10.1016/j.crhy.2013.09.013}}.
\newline\urlprefix\url{http://dx.doi.org/10.1016/j.crhy.2013.09.013}

\bibitem{Ryu2010}
S.~Ryu, A.~P. Schnyder, A.~Furusaki, A.~W.~W. Ludwig,
  \href{http://dx.doi.org/10.1088/1367-2630/12/6/065010}{Topological insulators
  and superconductors: tenfold way and dimensional hierarchy}, New J. Phys.
  12~(6) (2010) 065010.
\newblock \href {https://doi.org/10.1088/1367-2630/12/6/065010}
  {\path{doi:10.1088/1367-2630/12/6/065010}}.
\newline\urlprefix\url{http://dx.doi.org/10.1088/1367-2630/12/6/065010}

\bibitem{Qi2011}
X.-L. Qi, S.-C. Zhang,
  \href{http://dx.doi.org/10.1103/RevModPhys.83.1057}{Topological insulators
  and superconductors}, Rev. Mod. Phys. 83~(4) (2011) 1057–1110.
\newblock \href {https://doi.org/10.1103/revmodphys.83.1057}
  {\path{doi:10.1103/revmodphys.83.1057}}.
\newline\urlprefix\url{http://dx.doi.org/10.1103/RevModPhys.83.1057}

\bibitem{Ssstrunk2015}
R.~S\"{u}sstrunk, S.~D. Huber,
  \href{http://dx.doi.org/10.1126/science.aab0239}{Observation of phononic
  helical edge states in a mechanical topological insulator}, Science
  349~(6243) (2015) 47–50.
\newblock \href {https://doi.org/10.1126/science.aab0239}
  {\path{doi:10.1126/science.aab0239}}.
\newline\urlprefix\url{http://dx.doi.org/10.1126/science.aab0239}

\bibitem{Goldman2016}
N.~Goldman, J.~C. Budich, P.~Zoller,
  \href{http://dx.doi.org/10.1038/nphys3803}{Topological quantum matter with
  ultracold gases in optical lattices}, Nature Phys. 12~(7) (2016) 639–645.
\newblock \href {https://doi.org/10.1038/nphys3803}
  {\path{doi:10.1038/nphys3803}}.
\newline\urlprefix\url{http://dx.doi.org/10.1038/nphys3803}

\bibitem{Ozawa2019}
T.~Ozawa, H.~M. Price, A.~Amo, N.~Goldman, M.~Hafezi, L.~Lu, M.~C. Rechtsman,
  D.~Schuster, J.~Simon, O.~Zilberberg, I.~Carusotto,
  \href{http://dx.doi.org/10.1103/RevModPhys.91.015006}{Topological photonics},
  Rev. Mod. Phys. 91~(1) (Mar. 2019).
\newblock \href {https://doi.org/10.1103/revmodphys.91.015006}
  {\path{doi:10.1103/revmodphys.91.015006}}.
\newline\urlprefix\url{http://dx.doi.org/10.1103/RevModPhys.91.015006}

\bibitem{Malki2020}
M.~Malki, G.~S. Uhrig,
  \href{http://dx.doi.org/10.1209/0295-5075/132/20003}{Topological magnetic
  excitations}, EPL 132~(2) (2020) 20003.
\newblock \href {https://doi.org/10.1209/0295-5075/132/20003}
  {\path{doi:10.1209/0295-5075/132/20003}}.
\newline\urlprefix\url{http://dx.doi.org/10.1209/0295-5075/132/20003}

\bibitem{Ashida2020}
Y.~Ashida, Z.~Gong, M.~Ueda,
  \href{http://dx.doi.org/10.1080/00018732.2021.1876991}{Non-{H}ermitian
  physics}, Adv. Phys. 69~(3) (2020) 249–435.
\newblock \href {https://doi.org/10.1080/00018732.2021.1876991}
  {\path{doi:10.1080/00018732.2021.1876991}}.
\newline\urlprefix\url{http://dx.doi.org/10.1080/00018732.2021.1876991}

\bibitem{Tlusty2021}
T.~Tlusty, \href{http://dx.doi.org/10.1103/PhysRevE.104.025002}{Exceptional
  topology in ordinary soft matter}, Phys. Rev. E 104~(2) (Aug. 2021).
\newblock \href {https://doi.org/10.1103/physreve.104.025002}
  {\path{doi:10.1103/physreve.104.025002}}.
\newline\urlprefix\url{http://dx.doi.org/10.1103/PhysRevE.104.025002}

\bibitem{Hurst2022}
H.~M. Hurst, B.~Flebus,
  \href{http://dx.doi.org/10.1063/5.0124841}{Non-{H}ermitian physics in
  magnetic systems}, J. Appl. Phys. 132~(22) (Dec. 2022).
\newblock \href {https://doi.org/10.1063/5.0124841}
  {\path{doi:10.1063/5.0124841}}.
\newline\urlprefix\url{http://dx.doi.org/10.1063/5.0124841}

\bibitem{Banerjee2023}
A.~Banerjee, R.~Sarkar, S.~Dey, A.~Narayan,
  \href{http://dx.doi.org/10.1088/1361-648X/acd1cb}{Non-{H}ermitian topological
  phases: principles and prospects}, J. Phys.: Condens. Matter 35~(33) (2023)
  333001.
\newblock \href {https://doi.org/10.1088/1361-648x/acd1cb}
  {\path{doi:10.1088/1361-648x/acd1cb}}.
\newline\urlprefix\url{http://dx.doi.org/10.1088/1361-648X/acd1cb}

\bibitem{Yu2024}
T.~Yu, J.~Zou, B.~Zeng, J.~Rao, K.~Xia,
  \href{http://dx.doi.org/10.1016/j.physrep.2024.01.006}{Non-{H}ermitian
  topological magnonics}, Phys. Rep. 1062 (2024) 1–86.
\newblock \href {https://doi.org/10.1016/j.physrep.2024.01.006}
  {\path{doi:10.1016/j.physrep.2024.01.006}}.
\newline\urlprefix\url{http://dx.doi.org/10.1016/j.physrep.2024.01.006}

\bibitem{Yao2018}
S.~Yao, F.~Song, Z.~Wang,
  \href{http://dx.doi.org/10.1103/PhysRevLett.121.136802}{Non-{H}ermitian
  {C}hern bands}, Phy. Rev. Lett. 121~(13) (Sep. 2018).
\newblock \href {https://doi.org/10.1103/physrevlett.121.136802}
  {\path{doi:10.1103/physrevlett.121.136802}}.
\newline\urlprefix\url{http://dx.doi.org/10.1103/PhysRevLett.121.136802}

\bibitem{Hirschberger2015}
M.~Hirschberger, R.~Chisnell, Y.~S. Lee, N.~P. Ong,
  \href{https://link.aps.org/doi/10.1103/PhysRevLett.115.106603}{Thermal {H}all
  effect of spin excitations in a kagome magnet}, Phys. Rev. Lett. 115 (2015)
  106603.
\newblock \href {https://doi.org/10.1103/PhysRevLett.115.106603}
  {\path{doi:10.1103/PhysRevLett.115.106603}}.
\newline\urlprefix\url{https://link.aps.org/doi/10.1103/PhysRevLett.115.106603}

\bibitem{Chen2018_PRB98}
Y.~Chen, H.~Zhai, \href{http://dx.doi.org/10.1103/PhysRevB.98.245130}{{H}all
  conductance of a non-{H}ermitian {C}hern insulator}, Phys. Rev. B 98~(24)
  (Dec. 2018).
\newblock \href {https://doi.org/10.1103/physrevb.98.245130}
  {\path{doi:10.1103/physrevb.98.245130}}.
\newline\urlprefix\url{http://dx.doi.org/10.1103/PhysRevB.98.245130}

\bibitem{Hirsbrunner2019_PRB100}
M.~R. Hirsbrunner, T.~M. Philip, M.~J. Gilbert,
  \href{http://dx.doi.org/10.1103/PhysRevB.100.081104}{Topology and observables
  of the non-{H}ermitian {C}hern insulator}, Phys. Rev. B 100~(8) (Aug. 2019).
\newblock \href {https://doi.org/10.1103/physrevb.100.081104}
  {\path{doi:10.1103/physrevb.100.081104}}.
\newline\urlprefix\url{http://dx.doi.org/10.1103/PhysRevB.100.081104}

\bibitem{Canals2002}
B.~Canals, \href{http://dx.doi.org/10.1103/PhysRevB.65.184408}{From the square
  lattice to the checkerboard lattice: Spin-wave and large-n limit analysis},
  Phys. Rev. B 65~(18) (Apr. 2002).
\newblock \href {https://doi.org/10.1103/physrevb.65.184408}
  {\path{doi:10.1103/physrevb.65.184408}}.
\newline\urlprefix\url{http://dx.doi.org/10.1103/PhysRevB.65.184408}

\bibitem{Fujimoto2002}
S.~Fujimoto,
  \href{https://link.aps.org/doi/10.1103/PhysRevLett.89.226402}{Geometrical-frustration-induced
  (semi)metal-to-insulator transition}, Phys. Rev. Lett. 89 (2002) 226402.
\newblock \href {https://doi.org/10.1103/PhysRevLett.89.226402}
  {\path{doi:10.1103/PhysRevLett.89.226402}}.
\newline\urlprefix\url{https://link.aps.org/doi/10.1103/PhysRevLett.89.226402}

\bibitem{Bernier2004}
J.-S. Bernier, C.-H. Chung, Y.~B. Kim, S.~Sachdev,
  \href{http://dx.doi.org/10.1103/PhysRevB.69.214427}{Planar pyrochlore
  antiferromagnet: {A} large-${N}$ analysis}, Phys. Rev. B 69~(21) (Jun. 2004).
\newblock \href {https://doi.org/10.1103/physrevb.69.214427}
  {\path{doi:10.1103/physrevb.69.214427}}.
\newline\urlprefix\url{http://dx.doi.org/10.1103/PhysRevB.69.214427}

\bibitem{Pollmann2006}
F.~Pollmann, J.~J. Betouras, K.~Shtengel, P.~Fulde,
  \href{http://dx.doi.org/10.1103/PhysRevLett.97.170407}{Correlated fermions on
  a checkerboard lattice}, Phys. Rev. Lett. 97~(17) (Oct. 2006).
\newblock \href {https://doi.org/10.1103/physrevlett.97.170407}
  {\path{doi:10.1103/physrevlett.97.170407}}.
\newline\urlprefix\url{http://dx.doi.org/10.1103/PhysRevLett.97.170407}

\bibitem{Yoshioka2008}
T.~Yoshioka, A.~Koga, N.~Kawakami,
  \href{https://link.aps.org/doi/10.1103/PhysRevB.78.165113}{Frustration
  effects in an anisotropic checkerboard lattice {H}ubbard model}, Phys. Rev. B
  78 (2008) 165113.
\newblock \href {https://doi.org/10.1103/PhysRevB.78.165113}
  {\path{doi:10.1103/PhysRevB.78.165113}}.
\newline\urlprefix\url{https://link.aps.org/doi/10.1103/PhysRevB.78.165113}

\bibitem{Santos_2010}
E.~G. Santos, J.~R. Iglesias, C.~Lacroix, M.~A. Gusmão,
  \href{https://dx.doi.org/10.1088/0953-8984/22/21/215701}{A two-band model for
  superconductivity in the checkerboard lattice}, J. Phys.: Condens. Matter
  22~(21) (2010) 215701.
\newblock \href {https://doi.org/10.1088/0953-8984/22/21/215701}
  {\path{doi:10.1088/0953-8984/22/21/215701}}.
\newline\urlprefix\url{https://dx.doi.org/10.1088/0953-8984/22/21/215701}

\bibitem{Sun2009}
K.~Sun, H.~Yao, E.~Fradkin, S.~A. Kivelson,
  \href{https://link.aps.org/doi/10.1103/PhysRevLett.103.046811}{Topological
  insulators and nematic phases from spontaneous symmetry breaking in 2d fermi
  systems with a quadratic band crossing}, Phys. Rev. Lett. 103 (2009) 046811.
\newblock \href {https://doi.org/10.1103/PhysRevLett.103.046811}
  {\path{doi:10.1103/PhysRevLett.103.046811}}.
\newline\urlprefix\url{https://link.aps.org/doi/10.1103/PhysRevLett.103.046811}

\bibitem{Sun2011}
K.~Sun, Z.~Gu, H.~Katsura, S.~Das~Sarma,
  \href{http://dx.doi.org/10.1103/PhysRevLett.106.236803}{Nearly flatbands with
  nontrivial topology}, Phys. Rev. Lett. 106~(23) (Jun. 2011).
\newblock \href {https://doi.org/10.1103/physrevlett.106.236803}
  {\path{doi:10.1103/physrevlett.106.236803}}.
\newline\urlprefix\url{http://dx.doi.org/10.1103/PhysRevLett.106.236803}

\bibitem{Katsura2010_linegraph}
H.~Katsura, I.~Maruyama, A.~Tanaka, H.~Tasaki,
  \href{http://dx.doi.org/10.1209/0295-5075/91/57007}{Ferromagnetism in the
  {H}ubbard model with topological/non-topological flat bands}, Europhys. Lett.
  91~(5) (2010) 57007.
\newblock \href {https://doi.org/10.1209/0295-5075/91/57007}
  {\path{doi:10.1209/0295-5075/91/57007}}.
\newline\urlprefix\url{http://dx.doi.org/10.1209/0295-5075/91/57007}

\bibitem{Liu2022_linegraphs}
H.~Liu, G.~Sethi, S.~Meng, F.~Liu,
  \href{http://dx.doi.org/10.1103/PhysRevB.105.085128}{Orbital design of flat
  bands in non-line-graph lattices via line-graph wave functions}, Phys. Rev. B
  105~(8) (Feb. 2022).
\newblock \href {https://doi.org/10.1103/physrevb.105.085128}
  {\path{doi:10.1103/physrevb.105.085128}}.
\newline\urlprefix\url{http://dx.doi.org/10.1103/PhysRevB.105.085128}

\bibitem{Liu2017}
X.-P. Liu, Y.~Zhou, Y.-F. Wang, C.-D. Gong,
  \href{http://dx.doi.org/10.1088/1367-2630/aa8022}{Characterizations of
  topological superconductors: {C}hern numbers, edge states and {M}ajorana zero
  modes}, New J. Phys. 19~(9) (2017) 093018.
\newblock \href {https://doi.org/10.1088/1367-2630/aa8022}
  {\path{doi:10.1088/1367-2630/aa8022}}.
\newline\urlprefix\url{http://dx.doi.org/10.1088/1367-2630/aa8022}

\bibitem{Pires2019}
A.~Pires, \href{http://dx.doi.org/10.1016/j.physleta.2019.125887}{Magnon spin
  {N}ernst effect on the antiferromagnetic checkerboard lattice}, Phys. Lett. A
  383~(32) (2019) 125887.
\newblock \href {https://doi.org/10.1016/j.physleta.2019.125887}
  {\path{doi:10.1016/j.physleta.2019.125887}}.
\newline\urlprefix\url{http://dx.doi.org/10.1016/j.physleta.2019.125887}

\bibitem{Pires2020}
A.~S.~T. Pires, \href{https://doi.org/10.1016/j.physe.2019.113899}{Topological
  magnons in the antiferromagnetic checkerboard lattice}, Physica E 118 (2020)
  113899.
\newblock \href {https://doi.org/10.1016/j.physe.2019.113899}
  {\path{doi:10.1016/j.physe.2019.113899}}.
\newline\urlprefix\url{https://doi.org/10.1016/j.physe.2019.113899}

\bibitem{Pires2021_2}
A.~Pires, \href{http://dx.doi.org/10.1016/j.physb.2020.412490}{Topological
  magnons on the checkerboard lattice}, Physica B 602 (2021) 412490.
\newblock \href {https://doi.org/10.1016/j.physb.2020.412490}
  {\path{doi:10.1016/j.physb.2020.412490}}.
\newline\urlprefix\url{http://dx.doi.org/10.1016/j.physb.2020.412490}

\bibitem{Ma2020}
D.-S. Ma, Y.~Xu, C.~S. Chiu, N.~Regnault, A.~A. Houck, Z.~Song, B.~A. Bernevig,
  \href{https://link.aps.org/doi/10.1103/PhysRevLett.125.266403}{Spin-orbit-induced
  topological flat bands in line and split graphs of bipartite lattices}, Phys.
  Rev. Lett. 125 (2020) 266403.
\newblock \href {https://doi.org/10.1103/PhysRevLett.125.266403}
  {\path{doi:10.1103/PhysRevLett.125.266403}}.
\newline\urlprefix\url{https://link.aps.org/doi/10.1103/PhysRevLett.125.266403}

\bibitem{Zhang2021}
Z.~Zhang, W.~Feng, Y.~Yao, B.~Tang,
  \href{http://dx.doi.org/10.1016/j.physleta.2021.127630}{Photoinduced
  {F}loquet topological magnons in a ferromagnetic checkerboard lattice}, Phys.
  Lett. A 414 (2021) 127630.
\newblock \href {https://doi.org/10.1016/j.physleta.2021.127630}
  {\path{doi:10.1016/j.physleta.2021.127630}}.
\newline\urlprefix\url{http://dx.doi.org/10.1016/j.physleta.2021.127630}

\bibitem{Kabbour2008}
H.~Kabbour, E.~Janod, B.~Corraze, M.~Danot, C.~Lee, M.-H. Whangbo, L.~Cario,
  \href{http://dx.doi.org/10.1021/ja711139g}{Structure and magnetic properties
  of oxychalcogenides ${A}_2{F}_2{F}e_2{O}{Q}_2$ ({A} = {S}r, {B}a; {Q} = {S},
  {S}e) with ${F}e_2{O}$ square planar layers representing an antiferromagnetic
  checkerboard spin lattice}, J. Am. Chem. Soc. 130~(26) (2008) 8261–8270.
\newblock \href {https://doi.org/10.1021/ja711139g}
  {\path{doi:10.1021/ja711139g}}.
\newline\urlprefix\url{http://dx.doi.org/10.1021/ja711139g}

\bibitem{Hu2023}
X.~Hu, R.-W. Zhang, D.-S. Ma, Z.~Cai, D.~Geng, Z.~Sun, Q.~Zhao, J.~Gao,
  P.~Cheng, L.~Chen, K.~Wu, Y.~Yao, B.~Feng,
  \href{http://dx.doi.org/10.1021/acs.nanolett.3c01111}{Realization of a
  two-dimensional checkerboard lattice in monolayer ${C}u_2{N}$}, Nano Lett.
  23~(12) (2023) 5610–5616.
\newblock \href {https://doi.org/10.1021/acs.nanolett.3c01111}
  {\path{doi:10.1021/acs.nanolett.3c01111}}.
\newline\urlprefix\url{http://dx.doi.org/10.1021/acs.nanolett.3c01111}

\bibitem{Sufyan2024}
A.~Sufyan, M.~Sajjad, J.~Andreas~Larsson,
  \href{http://dx.doi.org/10.1016/j.apsusc.2024.159474}{Evaluating the
  potential of planar checkerboard lattice ${C}u_2{N}$ monolayer as anode
  material for lithium and sodium-ion batteries using first-principles
  methods}, Appl. Surf. Sci. 654 (2024) 159474.
\newblock \href {https://doi.org/10.1016/j.apsusc.2024.159474}
  {\path{doi:10.1016/j.apsusc.2024.159474}}.
\newline\urlprefix\url{http://dx.doi.org/10.1016/j.apsusc.2024.159474}

\bibitem{Midtgaard2019}
J.~M. Midtgaard, Z.~Wu, Y.~Chen,
  \href{http://dx.doi.org/10.1140/epjb/e2019-100393-5}{Constraints on the
  energy spectrum of non-{H}ermitian models in open environments}, Eur. Phys.
  J. B 92~(11) (Nov. 2019).
\newblock \href {https://doi.org/10.1140/epjb/e2019-100393-5}
  {\path{doi:10.1140/epjb/e2019-100393-5}}.
\newline\urlprefix\url{http://dx.doi.org/10.1140/epjb/e2019-100393-5}

\bibitem{Brody2013}
D.~C. Brody,
  \href{http://dx.doi.org/10.1088/1751-8113/47/3/035305}{Biorthogonal quantum
  mechanics}, J. Phys. A: Math. Theor. 47~(3) (2013) 035305.
\newblock \href {https://doi.org/10.1088/1751-8113/47/3/035305}
  {\path{doi:10.1088/1751-8113/47/3/035305}}.
\newline\urlprefix\url{http://dx.doi.org/10.1088/1751-8113/47/3/035305}

\bibitem{Groenendijk2021}
S.~Groenendijk, T.~L. Schmidt, T.~Meng,
  \href{http://dx.doi.org/10.1103/PhysRevResearch.3.023001}{Universal {H}all
  conductance scaling in non-{H}ermitian {C}hern insulators}, Phys. Rev. Res.
  3~(2) (Apr. 2021).
\newblock \href {https://doi.org/10.1103/physrevresearch.3.023001}
  {\path{doi:10.1103/physrevresearch.3.023001}}.
\newline\urlprefix\url{http://dx.doi.org/10.1103/PhysRevResearch.3.023001}

\bibitem{Ishikawa1986}
K.~Ishikawa, T.~Matsuyama, \href{http://dx.doi.org/10.1007/BF01410451}{Magnetic
  field induced multi-component ${QED}_3$ and quantum {H}all effect}, Z. Phys.
  C - Particles and Fields 33~(1) (1986) 41–45.
\newblock \href {https://doi.org/10.1007/bf01410451}
  {\path{doi:10.1007/bf01410451}}.
\newline\urlprefix\url{http://dx.doi.org/10.1007/BF01410451}

\bibitem{Ishikawa1987}
K.~Ishikawa, T.~Matsuyama,
  \href{http://dx.doi.org/10.1016/0550-3213(87)90160-X}{A microscopic theory of
  the quantum {H}all effect}, Nucl. Phys. B 280 (1987) 523–548.
\newblock \href {https://doi.org/10.1016/0550-3213(87)90160-x}
  {\path{doi:10.1016/0550-3213(87)90160-x}}.
\newline\urlprefix\url{http://dx.doi.org/10.1016/0550-3213(87)90160-X}

\bibitem{TKNN}
D.~J. Thouless, M.~Kohmoto, M.~P. Nightingale, M.~den Nijs,
  \href{https://doi.org/10.1103/physrevlett.49.405}{Quantized {H}all
  conductance in a two-dimensional periodic potential}, Phys. Rev. Lett. 49~(6)
  (1982) 405--408.
\newblock \href {https://doi.org/10.1103/physrevlett.49.405}
  {\path{doi:10.1103/physrevlett.49.405}}.
\newline\urlprefix\url{https://doi.org/10.1103/physrevlett.49.405}

\bibitem{Pires_topology}
A.~S.~T. Pires, A brief introduction to topology and differential geometry in
  condensed matter physics, IOP Concise Physics, Morgan \& Claypool, San
  Rafael, CA, 2019.

\bibitem{Shen2018}
H.~Shen, B.~Zhen, L.~Fu,
  \href{http://dx.doi.org/10.1103/PhysRevLett.120.146402}{Topological band
  theory for non-{H}ermitian {H}amiltonians}, Phys. Rev. Lett. 120~(14) (Apr.
  2018).
\newblock \href {https://doi.org/10.1103/physrevlett.120.146402}
  {\path{doi:10.1103/physrevlett.120.146402}}.
\newline\urlprefix\url{http://dx.doi.org/10.1103/PhysRevLett.120.146402}

\bibitem{Zhang2018_HEP}
X.-L. Zhang, C.~T. Chan,
  \href{http://dx.doi.org/10.1103/PhysRevA.98.033810}{Hybrid exceptional point
  and its dynamical encircling in a two-state system}, Phys. Rev. A 98~(3)
  (Sep. 2018).
\newblock \href {https://doi.org/10.1103/physreva.98.033810}
  {\path{doi:10.1103/physreva.98.033810}}.
\newline\urlprefix\url{http://dx.doi.org/10.1103/PhysRevA.98.033810}

\bibitem{Jin2020}
L.~Jin, H.~C. Wu, B.-B. Wei, Z.~Song,
  \href{http://dx.doi.org/10.1103/PhysRevB.101.045130}{Hybrid exceptional point
  created from type-iii {D}irac point}, Phys. Rev. B 101~(4) (Jan. 2020).
\newblock \href {https://doi.org/10.1103/physrevb.101.045130}
  {\path{doi:10.1103/physrevb.101.045130}}.
\newline\urlprefix\url{http://dx.doi.org/10.1103/PhysRevB.101.045130}

\bibitem{Zhou2018}
H.~Zhou, C.~Peng, Y.~Yoon, C.~W. Hsu, K.~A. Nelson, L.~Fu, J.~D. Joannopoulos,
  M.~Soljačić, B.~Zhen,
  \href{http://dx.doi.org/10.1126/science.aap9859}{Observation of bulk {F}ermi
  arc and polarization half charge from paired exceptional points}, Science
  359~(6379) (2018) 1009–1012.
\newblock \href {https://doi.org/10.1126/science.aap9859}
  {\path{doi:10.1126/science.aap9859}}.
\newline\urlprefix\url{http://dx.doi.org/10.1126/science.aap9859}

\bibitem{Dey2024}
S.~Dey, A.~Banerjee, D.~Chowdhury, A.~Narayan,
  \href{http://dx.doi.org/10.1088/1367-2630/ad2b0e}{{H}all conductance of a
  non-{H}ermitian {W}eyl semimetal}, New J. Phys. 26~(2) (2024) 023057.
\newblock \href {https://doi.org/10.1088/1367-2630/ad2b0e}
  {\path{doi:10.1088/1367-2630/ad2b0e}}.
\newline\urlprefix\url{http://dx.doi.org/10.1088/1367-2630/ad2b0e}

\end{thebibliography}

\end{document}